\documentclass[%
 aip,
rsi,%
 amsmath,amssymb,
 reprint,%
]{revtex4-1}

\usepackage{graphicx}% Include figure files
\usepackage{subfig}
\usepackage{amssymb}
\usepackage{float}
\usepackage{todonotes}
\usepackage{changes}
 \usepackage{color}
\usepackage{gensymb}

\usepackage{amsmath}

\usepackage{algpseudocode,caption}
\usepackage[english]{babel}
\usepackage[utf8]{inputenc}
\usepackage{algorithm}

\usepackage{bm}% bold math
\DeclareMathOperator{\sech}{sech}

\begin{document}

\preprint{AIP/123-QED}
\title[Ocean bathymetry reconstruction from surface data using hydraulics theory]{Ocean bathymetry reconstruction from surface data using hydraulics theory}% Force line breaks with \\
\author{Subhajit Kar}

% \altaffiliation[Also at ]{Physics Department, XYZ University.}%Lines break automatically or can be forced with \\

%\collaboration{MUSO Collaboration}%\noaffiliation
\author{Anirban Guha}
 \email{anirbanguha.ubc@gmail.com}
\affiliation{
Environmental and Geophysical Fluids Group, Department of Mechanical Engineering, Indian Institute of Technology Kanpur, U.P. 208016, India.\\}%

\date{\today}% It is always \today, today,
             %  but any date may be explicitly specified

\begin{abstract}
 Here we propose a technique that successfully reconstructs ocean bathymetry from the free surface velocity and elevation data. This technique is based on the principles of
 open-channel hydraulics, according to which a sub-critical flow over a seamount creates a free surface dip. The proposed method recognizes that such free surface dip contains the signature of the bottom topography,  hence inverts the free surface to reconstruct the topography 
 accurately.
 %To test the applicability of this method in realistic situations, we simulated the Mediterranean sea using Massachusetts Institute of Technology general circulation model (MITgcm), and initialized it with re-analysis data.
 {
We applied our inversion technique on re-analysis data, and  reconstructed the Mediterranean and the Red sea bathymetries of  $1/12\degree$ resolution with approximately  $90$\% accuracy.}

 %We expect that our reconstruction technique, in conjunction with ship echo-sounding, will be able to provide a high resolution and accurate global bathymetry map in near future.   

%{Things remaining- (i) Abstract (ii) Expand introduction (iii) Finalize names of section (iv) Section 3 (v) Conclusion (vi) Change figures}
% \begin{description}
% %\item[Usage]
% %Secondary publications and information retrieval purposes.
 %\item[PACS numbers]
% \pacs{47.52.+j}
%\pacs{47.52.+j,47.35.Bb}
% %May be entered using the \verb+\pacs{#1}+ command.
% %\item[Structure]
% %You may use the \texttt{description} environment to structure your abstract;
% %use the optional argument of the \verb+\item+ command to give the category of each item. 
% \end{description}
\end{abstract}

% PACS, the Physics and Astronomy
                             % Classification Scheme.
%Use showkeys class option if keyword
                              %display desired

%\pacs{47.52.+j, 47.15.St, 05.45.Xt, 47.20.Ky}

% PACS, the Physics and Astronomy
% PACS, the Physics and Astronomy
                             % Classification Scheme.
%\keywords{Suggested keywords}%Use showkeys class option if keyword
                              %display desired
\maketitle

%\section{Introduction}
The ocean floor displays diverse geological features, such as seamounts, plateaus and other structures associated with intraplate volcanism \citep{smith1997global,wessel1997sizes}, subduction zones that can generate earthquakes and tsunamis \citep{mofjeld2004tsumani}, as well as regions rich in oil and gas \citep{fairhead2001satellite}. Detailed knowledge of ocean bathymetry is essential for understanding ocean circulation and mixing, which in turn moderates the earth's climate \citep{munk1998abyssal}.
%Seafloor topography influences the upwelling of nutrient rich water, which strongly affects marine biology \citep{ryan2009global,becker2009global}.
%Detailed bathymetric information is also important for submarine navigation, coastal resource management, placement of offshore platforms and pipelines, and management of marine fisheries \citep{ryan2009global}.
Bathymetry mapping is arguably one of the most important and challenging problems in oceanography \citep{nicholls2009detection}. Usually, ships equipped with echo sounders are deployed for the acquisition of high-resolution seafloor map. This process is difficult, expensive, and slow. It may cost billions of dollars, and respectively take 120 and 750 ship-years of survey time for mapping the deep and shallow oceans \citep{becker2009global}.
 Even after five decades of ship-based surveying, $90$\% (at $1$ minute resolution) of the global seafloor is still unexplored.

 While ship echo-sounding directly maps the ocean floor, satellite altimetry provides an indirect approach to bathymetry reconstruction. Currently, the only available altimetry based bathymetry reconstruction technique, the ``altimetric bathymetry'',  provides lower resolution and accuracy than ship-based mapping  \citep{becker2009global,smith2004conventional}. 
  The underlying principle of altimetric bathymetry is the following: seamounts add extra pull to the earth's gravitational field  and therefore draws more seawater around them, which leads to a small outward bulge of the marine geoid  \citep{smith2004conventional}. The seafloor can thus be reconstructed by analysing such minute dips and bulges of the geoid profile.
This principle is expected to work in the $\sim 15$ -- $160$ km wavelength band where marine gravity anomaly and seafloor topography are highly correlated \citep{smith1994bathymetric}. 

\begin{figure*} 
\centering
\includegraphics[angle=0,width=0.7\linewidth]{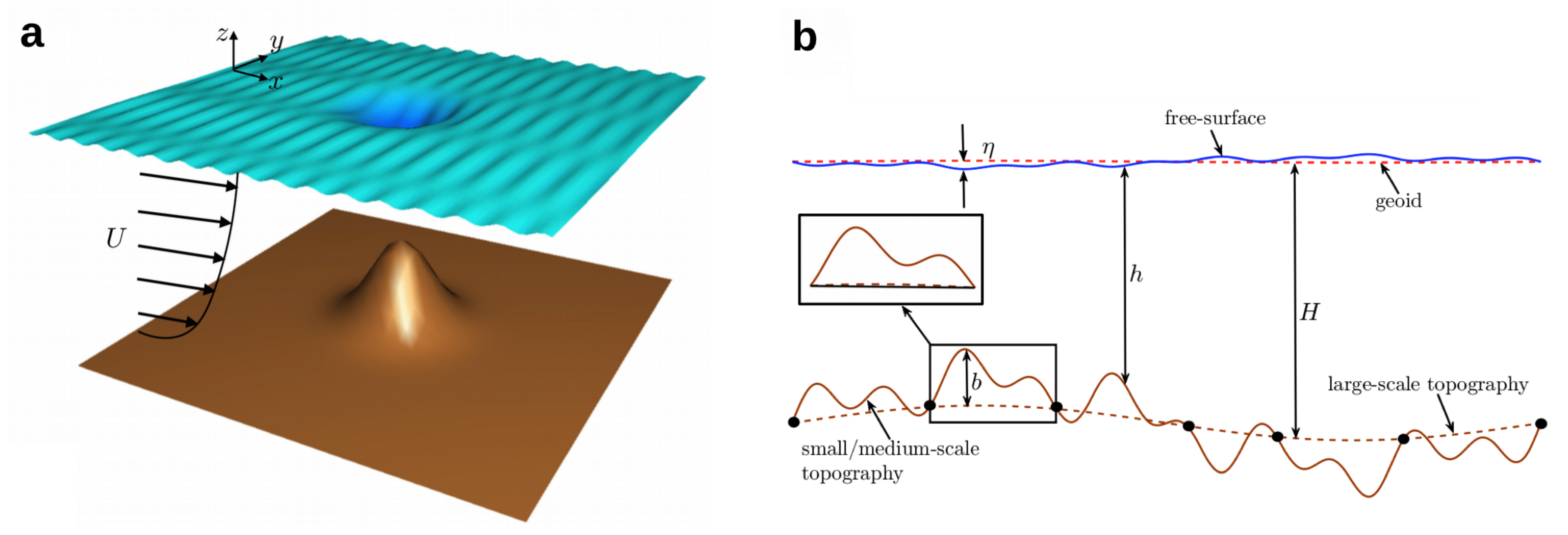}
\caption{(a) Schematic of a unidirectional sub-critical flow showing that the free surface (exaggerated) dips down while flowing over a seamount. This permanent feature at the free surface is present along with transient features like surface gravity waves. (b) Medium/small-scale topographic features (of height $b$) present on the top of large-scale features. The free surface elevation $\eta$ and the `mean depth' $H$ are calculated with respect to the the geoid, while the water depth $h$ is the distance between the free surface and the sea-bed. Between two successive black dots, the large-scale topography is nearly flat (see inset). }
\label{fig:one}
\end{figure*}

Attempts have also been made to reconstruct ocean bathymetry using the principles of fluid dynamics \cite{vasan2013inverse}. 
Vasan and Deconinck \cite{vasan2013inverse} emphasize the ill-posed nature of this inverse problem, and show that bathymetry reconstruction is possible in idealized scenarios and under certain regimes, specifically the shallow water regime. They found that for bathymetry  reconstruction, the surface elevation and
its first two time derivatives as functions of the horizontal variable at several successive instances of time are needed. The practical feasibility of obtaining the input data, and hence the application of this method in real-world scenario is questionable.\\
\,\,Here we propose a new inversion technique that reconstructs  bottom topography with a uniform resolution and  reasonably high accuracy  from the free surface elevation and velocity field.  
Since both ocean surface elevation and velocity data can be obtained from satellite altimetry, our proposed  technique can be directly implemented to reconstruct real ocean bathymetry.\\
\,\,Large scale oceanic flows are in geostrophic and hydrostatic balance, which cause the free surface to  tilt permanently \citep{vallis2017atmospheric}.  
 Semi-permanent free surface tilts are also produced by wind-stress and flow over topography. %an example of the latter is the  surface dip above the Charleston bump -- a seamount located off the South Carolina coast \citep{xie2007effect}. 
In the latter case, the underlying principle can be explained using the theory of \textit{open-channel hydraulics};  see Fig.\ \ref{fig:one}(a). 
Oceanic circulation is strongly affected by its geometric shallowness. This significantly simplifies the governing equations of motion (vertical dynamics become negligible in comparison to the horizontal), yielding the celebrated shallow water equations (SWEs) \citep{vallis2017atmospheric}, which form the basis of open-channel hydraulics. 
 In presence of planetary rotation and  absence of viscous forces, the two-dimensional (2D) SWEs in Cartesian coordinates are given by
\begin{align} 
  \label{eqn1}
\frac {\partial h}{\partial t} + \
\frac {\partial (uh)}{\partial x} + 
\frac {\partial (vh)}{\partial y}  = 0,  \\
  \label{eqn2}
\frac {\partial u}{\partial t} + 
u \frac {\partial u}{\partial x} + 
v \frac {\partial u}{\partial y} - f v = 
-g\frac {\partial \eta}{\partial x}, \\
  \label{eqn3}
\frac {\partial v}{\partial t} + 
u \frac {\partial v}{\partial x} + 
v \frac {\partial v}{\partial y} + f u = 
-g\frac {\partial \eta}{\partial y}.
\end{align} 
Here $h(x,y,t)$ is the water depth, $u(x,y,t)$ and $v(x,y,t)$ are respectively the $x$ (zonal) and $y$ (meridional) components of the horizontal velocity,  
 $f$ is the Coriolis frequency ($f\equiv 2\Omega \sin \theta$, where $\Omega=7.2921 \times 10^{-5}$ s$^{-1}$ is Earth's rotation rate and $\theta$ is the latitude of interest), $g=9.81$ ms$^{-2}$ is the acceleration due to gravity, 
 %$\rho_w$ is the density of the water ($998.2$ kgm$^{-3}$),
%  \begin{equation}
%  \label{eq:for_wind_stress}
%   \bm{\tau}=\,(\tau_x,\,\tau_y)=\rho_a C_d \mathbf{u}_{a} |\mathbf {u}_{a}|
%  \end{equation}
% is  the wind-stress (following \citet{gill1982atmosphere}, where $\rho_a=1.2$ kg/m$^3$ is the density of air, $C_d$ is the drag coefficient and $\mathbf{u}_{a}$ is the wind velocity),  
$\eta(x,y,t)=h(x,y,t)+b(x,y)-H$ is the free surface elevation, $H$ and $b$ respectively being the mean depth and the bottom topography; see Fig.\ \ref{fig:one}(b).

%When the wind speed is low, wind-stress has a small effect on the free surface shape, hence we can simplify the wind-stress terms $\tau_{x}/(\rho_w h)$ and $\tau_{y}/(\rho_w h)$ in  Eq.\ (\ref{eqn2})--Eq.\ (\ref{eqn3}) by approximating $h$ by $H$ (the error produced by this approximation is negligible if $b/H$ is small, which holds reasonably well for oceans). 

For a steady, one-dimensional (1D) flow in the absence of rotation,  Eqs.\ (\ref{eqn1})-(\ref{eqn3}) can be highly simplified.
These equations under linearization about the base velocity $U$ and the base height $H$ 
 yield \citep{whitehead1998topographic,henderson1996open,chaudhry2007open}
\begin{equation}
\frac{db}{dx}=\left(\frac{Fr^2-1}{Fr^2}\right)\frac{d\eta}{dx},
\label{s4}
\end{equation}
where $Fr \equiv U/\sqrt{gH}$ denotes the Froude number. For sub-critical flows $Fr<1$, hence the bottom slope $db/dx$ and the free surface slope $d\eta/dx$ have opposite signs. This  mathematically justifies why flow over a bump produces a free surface dip.   The concept of open-channel flows can be extended to oceans. Oceanic flows are usually highly sub-critical since $U \sim O(0.1 - 1)$ ms$^{-1}$, while $c \approx 200$ ms$^{-1}$ for an ocean with  $H=4$ km. Hence one can expect a small depression at the ocean free surface right above a seamount.

Fourier transform of  Eq.\ (\ref{s4}) relates the  amplitude of the free surface dip, $\hat{\eta}$, to the  topography  amplitude, $\hat{b}$:
\begin{equation}
\label{s5}
 \hat{\eta}(k)=\left(\frac{Fr^2}{Fr^2-1 }\right)\hat{b}(k),
\end{equation}
where  $k$ denotes the  wavenumber and `hat' denotes the transformed variable (signifying the amplitude corresponding to $k$). Since in oceans   $Fr \sim 0.01$ -- $0.001$, the free surface imprint of a topography %{with horizontal scale of $10$ km with} 
$\hat{b}=100$ m will be  $\sim 10 - 0.1$ mm.  Modern altimeters have the ability to  largely detect such small amplitude free surface anomalies\citep{smith2004conventional}. 

Based on the fundamental theory of open-channel hydraulics we make two crucial observations:  (i) whenever there is a quasi-steady \textit{open flow} over a topography, the shape of the latter gets \textit{imprinted} on the free surface, and (ii) the imprint is  \textit{quasi-permanent}, and can therefore be inverted to reconstruct the bottom topography.

As we have already shown, in an idealized, steady 1D flow, the bottom topography can be successfully reconstructed from the free surface elevation using Eq.\ (\ref{s5}). In a real ocean scenario,   the free surface elevation contains transient features like surface waves along with  the following major quasi-permanent features: (i) the tilt due to the geostrophic flow, $\eta_g$, (ii) tilt due to wind stress, $\eta_s$ and (iii) topography's free surface imprint, $\eta_b$. For now we will assume that there are no wind-stresses, hence $\eta_s=0$. If the geostrophic velocity field $\bf{u_g}$ is known, $\eta_g$ can be computed as follows:
 \begin{equation}
\nabla \eta_g=-\frac{f}{g}\mathbf{\hat{k}}\times \mathbf{u_g},
 \label{eq_geo_tilt}
 \end{equation}
 where $\bf{\hat{k}}$ is the unit-vector in the vertical direction. 
{Following Vallis \cite{vallis2017atmospheric}, the non-dimensional form of Eqs.\ (2)--(3) can be written as:
\begin{align} 
  \label{eqn123}
Ro \left[ \frac{\partial \hat{\mathbf{u}}}{\partial \hat{t}} + (\hat{\mathbf{u}} \cdot \nabla) \hat{\mathbf{u}} \right] + \hat{z} \times \hat{\mathbf{u}} = \frac{Ro}{{Fr}^2} \lambda \nabla \hat{\eta}.
\end{align} 
Here, $Ro = U/{fL}$ is the Rossby number, $L$ is the horizontal length scale and $L/U$ is the advective timescale; $\lambda=\Delta\eta$/$H$, $h=H(1+\lambda\hat{\eta})-b$, $\hat{\eta}=\eta/\Delta\eta$, and $\Delta\eta$ is the scale of $\eta$. Variables with `hat' denote the non-dimensional variables. Note that $Fr$ is independent of $L$ and  is usually small in oceans ($Fr \approx 0.01$ -- $0.001$). Since $Ro$ can change depending on $L$, we choose $Fr$ as the `small parameter' and vary $Ro$.\\ 
\indent When  $L\approx 1000$ km, $Ro$ is a small number. The choice $Fr\sim Ro \sim \epsilon$, where $0<\epsilon \ll O(1)$ is a small parameter, leads to the balance between the Coriolis term and the RHS in Eq.\ (\ref{eqn123}), and for this  we must have $$\frac{Ro}{{Fr}^2} \lambda \sim O(1). $$ This is nothing but the geostrophic balance, i.e.\ Eq.\ (\ref{eq_geo_tilt}). Since $\lambda\sim\epsilon$, we observe that  $$\Delta\eta_{g}\sim\epsilon H.$$
\indent When $L \ll 100$ km, i.e.\ typical bathymetry scales we are interested in reconstructing, we find that $Ro \gtrsim O(1)$ (rotation plays a minor role). Hence the balance yields $\lambda \sim Fr^2\sim \epsilon^2$, which straightforwardly implies
% However $Fr$ is still a small number ($\epsilon$), and for the right hand side to balance the left, again we must have $$\frac{Ro}{{Fr}^2} \lambda = O(1).$$
% Accordingly, $\lambda=\epsilon^2$, and hence 
$$\Delta\eta_{b}=\epsilon^2 H.$$
Thus $\eta_g \gg \eta_b$, which means that the time average of the free surface elevation (by which transient features are removed)  $\eta$ can be expressed as a two-term perturbation expansion
\begin{equation}
\langle \eta \rangle=\eta_g+\eta_b,
\label{eq:eta_avg}
\end{equation}
 where the angle brackets denote time averaging. }
Once $\eta_g$ is removed from the free surface by applying Eq.\ (\ref{eq_geo_tilt}), the only free surface feature left that would be left is $\eta_b$.  

%  \begin{figure}
% \centering
% \includegraphics[width=1.0\textwidth]{wind}
% \caption{(a) Free surface elevation due to the combined effects of the bottom topography induced and the wind-stress induced tilts ($\eta_b+\eta_s$). (b) Structure of the wind-stress induced tilt $\eta_s$. (c) Imprint of the bottom topography on the free surface, $\eta_b$, obtained after removing the wind-stress induced tilt from the free surface.  The units of $x$ and $y$-coordinates  are in km. All the sub-figures share a common colorbar, and unit of  elevation is in m. }
% \label{fig:wind}
% \end{figure}

\begin{figure}
      \includegraphics[width=0.92\linewidth]{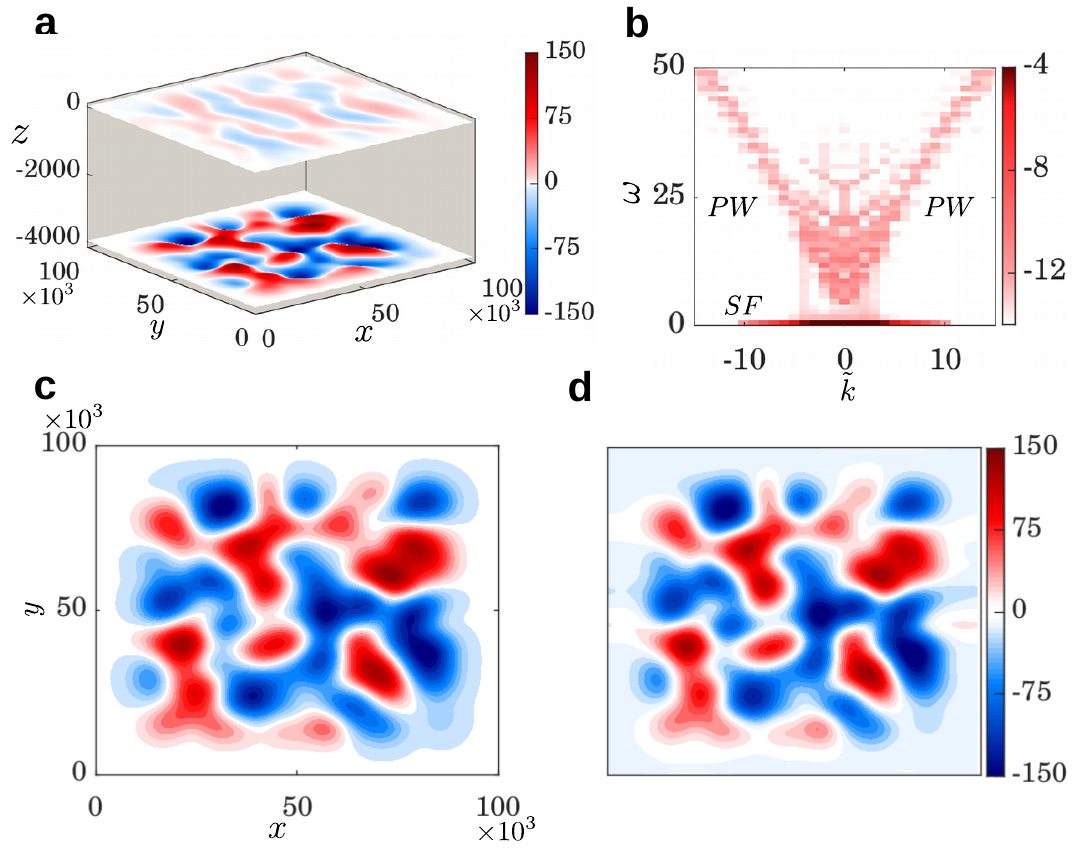}
\caption{(a) Imprint of the bottom topography on the free surface for 
      $Fr=0.001$. The free surface anomaly field (geostrophic effects removed) has been multiplied by
      $10^4$ to make it visible within the colorbar scale.  
      (b) Wavenumber ($\tilde{k}$, in km$^{-1}$) -- frequency ($\omega$, in s$^{-1}$) spectrum of the free surface anomaly.  `PW' denotes the dispersion relation of Poincar$\acute{\mathrm{e}}$ waves, while `SF' denotes the same for the `stationary features'. The colors denote magnitude (in log scale) of the free surface anomaly spectra.  (c) Actual topography, $b(x,y)$. (d) Topography reconstructed from the free surface data. For (a), (c) and (d), colors denote the height field (in m).}
\label{fig:print_2}
\end{figure}
Time averaging of the shallow water mass conservation equation, i.e.\  Eq.\ (\ref{eqn1}), and removal of $\eta_g$ from the free surface elevation yields
% The bathymetry induced  free surface tilt can be recovered by subtracting the geostrophic and wind-stress induced tilts from the total free surface elevation (see equation Eq.\ (\ref{eq:free_surf_elev_3})).
% The time-averaged  mass-conservation equation in Cartesian coordinates is given by
%\begin{equation}
%\label{SWC2}
%\frac{\partial}{\partial x} \bigl\langle  b u \bigr\rangle  + 
%\frac{\partial}{\partial y} \bigl\langle b v \bigr\rangle 
%= \frac{\partial}{\partial x} \bigl\langle  (\eta_b + H) u \bigr\rangle  + 
%\frac{\partial}{\partial y} \bigl\langle (\eta_b + H) v \bigr\rangle.
%\end{equation}
\begin{equation}
\label{SWC2}
\frac{\partial}{\partial x} (b \bigl\langle  u \bigr\rangle)  + 
\frac{\partial}{\partial y} (b \bigl\langle v \bigr\rangle) 
= \frac{\partial}{\partial x} \bigl\langle  (\eta_b + H) u \bigr\rangle  + 
\frac{\partial}{\partial y} \bigl\langle (\eta_b + H) v \bigr\rangle.
\end{equation}
After specifying appropriate boundary conditions for $b$ (zero at the boundaries), the above equation is solved using finite difference scheme to reconstruct $b$ \emph{entirely} from the free surface data ($u$, $v$ and $\eta_b$). Although $H$ is not a surface variable,   it is already known \textit{a-priori} from the coarse-resolution data. Since the free surface velocities and elevation can be obtained from satellite altimetry data, \ Eq.\ (\ref{SWC2}) can be directly used to reconstruct  ocean bathymetry.

\begin{figure*}
\centering
\includegraphics[width=0.9\textwidth]{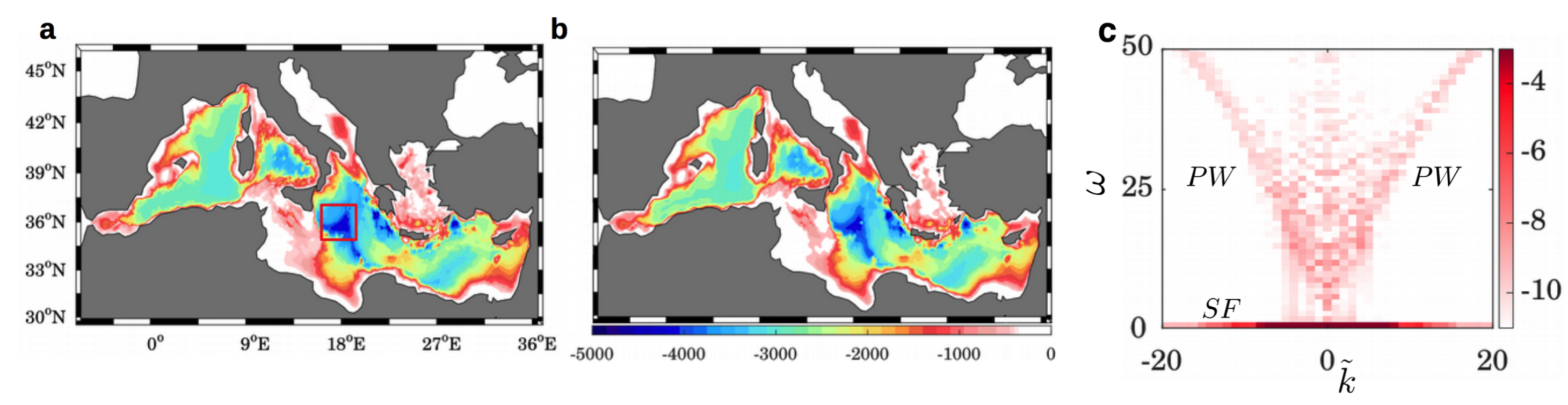}
\caption{Mediterranean sea bathymetry reconstruction using MITgcm. (a)  Actual bathymetry  from GEBCO, and (b) reconstructed bathymetry. The color contours represent  
depth $h$ (in m) from the free surface. (c) Wavenumber ($\tilde{k}$, in km$^{-1}$) -- frequency ($\omega$, in s$^{-1}$) spectrum of the free surface anomaly of  a part of the Mediterranean sea (marked by red-color box in
 (a)). The colors denote magnitude (in log scale) of the free surface anomaly spectra.}
\label{fig:sea}
\end{figure*}

% \begin{figure*}
% \centering
% \includegraphics[width=0.7\textwidth]{topo_mitgcm_1a_nw}
% \caption{Mediterranean sea bathymetry reconstruction using MITgcm. (a)  Actual bathymetry obtained from GEBCO, and (b) reconstructed bathymetry. The color contours represent  
% depth $h$ (in m) from the free surface.}
% \label{fig:sea}
% \end{figure*}

% \begin{figure}
% \includegraphics[width=0.70\linewidth]{meddi_igw}
% \caption{Wavenumber ($\tilde{k}$, in km$^{-1}$) -- frequency ($\omega$, in s$^{-1}$) spectrum of the free surface anomaly of part of the Mediterranean sea shown by a red-color box as shown in 
% Fig.\ \ref{fig:sea}. The colors denote magnitude (in log scale) of the free surface anomaly spectra.}
% \label{fig:igw}
% \end{figure}
%\section{Results} 
%\label{sec_results}
%\subsection{Bathymetry reconstruction of a \textit{toy ocean}}
First we consider  a simplified \textit{toy} ocean model that is governed by the 2D SWEs with planetary rotation, i.e.\  Eqs.\ (\ref{eqn1})-(\ref{eqn3}). The mean topography is a flat horizontal surface on which  Gaussian mountains and valleys of random amplitudes are added. 
The initial velocity field is under geostrophic and hydrostatic balance. We prescribe the initial height field as $H_0=H+\eta_g$, where the mean depth $H=4$ km, and the geostrophic tilt is
\begin{equation*}
\eta_g =  0.1 \tanh(\mathcal{Y})  
  + 0.03 \sech^2(\mathcal{Y}) \sin \Big(\frac{2 \pi x}{L_x} \Big),
\end{equation*}
% $L_x$ and $L_y$ respectively being the streamwise and spanwise extents. The corresponding geostrophic velocity scale yields  $Fr = 0.001$,  consistent with realistic parameters.  The wind-stress (m/s$^2$) is prescribed as  
% \begin{align*}
% %\label{eqn_wsH}
% %\tau_x = \frac{\partial \Phi}{\partial x} + \frac{\partial \Psi}{\partial y}  ,\,\, 
% %\tau_y = \frac{\partial \Phi}{\partial y} - \frac{\partial \Psi}{\partial x}.   
% \tau_x = 800 \pi \cos \Big(\frac{2\pi x}{L_x} \Big) \Big [ \frac{1}{L_x} \sin \Big(\frac{2 \pi y}{L_y} \Big) + \frac{1}{L_y} \cos \Big(\frac{2 \pi y}{L_y} \Big)  \Big], \\
% \tau_y = 800 \pi \sin \Big(\frac{2 \pi x}{L_x} \Big) \Big [ \frac{1}{L_y} \cos \Big(\frac{2 \pi y}{L_y} \Big) + \frac{1}{L_x} \sin \Big(\frac{2 \pi y}{L_y} \Big)  \Big].
% \end{align*}
where $\mathcal{Y}\equiv (0.5 L_y - y)/(2 L_y)$.
For numerical computation, a doubly-periodic horizontal domain of $L_x \times L_y = 10^5$ m $\times 10^5$ m  is assumed. The grid-size is $10^3$ m in both $x$ and $y$ directions, and time-step size is $1$ s. The numerical model uses second order central differencing for spatial and fourth order Runge-Kutta for temporal discretization, and is integrated for 10 days, by which a quasi-steady state is reached.  On time-averaging the free surface elevation using Eq.\ (\ref{eq:eta_avg}), we obtain the quasi-stationary features. The geostrophy induced tilt $\eta_g$  is removed using Eq.\ (\ref{eq_geo_tilt}). The remaining feature  contains the bathymetry induced tilt $\eta_b$. This $\eta_b$, also shown in 
Fig.\ \ref{fig:print_2}(a) (in Fig.\ \ref{fig:print_2}(b), it is shown as `SF' in the Fourier space),
%(e.g.\ rotating shallow water gravity waves or  `Poincar$\acute{\mathrm{e}}$ waves', indicated by `PW' in the dispersion diagram Fig.\ \ref{fig:print_2}(b)) 
is inverted to reconstruct the bottom topography using Eq.\ (\ref{SWC2}). 
The comparison between the actual  and  the reconstructed topography is shown in Figs.\ \ref{fig:print_2}(c)--\ref{fig:print_2}(d), the $L_2$-norm error is found to be 0.35\%. 
%The remaining feature, shown in Fig.\ \ref{fig:wind}A, contains both bathymetry induced tilt $\eta_b$ and  wind-stress induced tilt $\eta_s$. From the methodology outlined in  \ref{SST3}, first $\eta_s$ is evaluated using Eq.\ Eq.\ (\ref{eq:wind_stress_main})  (see Fig.\ \ref{fig:wind}B), and then  subtracted from the free surface elevation, yielding $\eta_b$ (Fig.\ \ref{fig:wind}C). Using this $\eta_b$ the bottom topography is reconstructed, 

The problem can also be  approached by performing Fourier-transform on the free surface anomaly data to obtain the wavenumber ($\tilde{k}$) -- frequency ($\omega$) spectrum ($\tilde{k}=\sqrt{k^2+l^2}$ is the magnitude of the horizontal wavenumber vector ($k,l$)), see  Fig.\ \ref{fig:print_2}(b). The spectrum shows both positively and negatively traveling Poincar$\acute{\mathrm{e}}$ waves (indicated by `PW'), whose dispersion relation is 
\begin{equation}
\omega^2= f^2 + {gH}{\tilde{k}}^2.
\label{eq:disp_rel}
\end{equation}
The stationary feature or `SF', located along $\omega \approx 0$, has the highest magnitude. Inverse Fourier transform of SF yields $\langle \eta \rangle$, and thus $\eta_b$, from which the bottom topography can be reconstructed using Eq.\ (\ref{SWC2}). An important point worth noting is that knowing $H$ \emph{a-priori} is not mandatory; the ($\omega$, $\tilde{k}$) values in Fig.\ \ref{fig:print_2}(b) can be substituted in   the  dispersion relation  Eq.\ (\ref{eq:disp_rel})  to obtain $H$.

%\subsection{Bathymetry reconstruction of a \textit{semi-realistic ocean} using MITgcm}
%\label{sec:MITgcm_1}
Based on the fundamental understanding of the 2D shallow water system, we have pursued bathymetry reconstruction of a more complicated, semi-realistic system. We have performed 
this particular exercise  keeping in mind that in real ocean scenario, the density changes are significantly small (approximately $\lesssim 2\%$ from a reference value). Furthermore, the large-scale motions are approximately in hydrostatic balance, hence the dynamics can be well explained using a simplified one-layer shallow water model \citep{gill1982atmosphere}.  In this regard we solve the 3D Navier-Stokes equations along with the evolution equations of  temperature and salinity using MITgcm. The latter is an open-source code that solves the following  non-linear, non-hydrostatic, primitive equations (under Boussinesq approximation)  in spherical coordinate system 
%$(r, \theta, \phi$ where, $r$ is the radius, $\theta$ is the latitude and $\phi$ is the longitude) 
using the finite volume method\citep{marshall1997fvm}.
%The following primitive equations, , have been solved:
% \begin{eqnarray}
% \label{gcm1}
% \frac{D \mathbf{u}}{Dt} + 2 \bm{\Omega} \times \mathbf{u} = -\frac{1}{\rho_w} \nabla p - \frac{\rho}{\rho_w}g  \bm{r} + \mathbf{F}, \\
% \label{gcm2}
% \nabla \cdot \mathbf{u} = 0, \\
% \label{gcm3}
% \rho = \rho(T, S), \\
% \label{gcm4}
% \frac{D T}{Dt} = Q_T, \\
% \label{gcm5}
% \frac{D S}{Dt} = Q_S.
% \end{eqnarray}
% Here $D/Dt\equiv\partial/\partial t + \mathbf{u} \cdot \nabla$ represents the material derivative and  $\mathbf{u}\equiv(u_r, u_\theta, u_\phi)$ is the velocity vector (the respective components being  radial, meridional and zonal).  The unit vector in the radial direction is denoted by $\bm{r}$. The quantities $\rho_w$, $T$ and $S$  respectively represent  reference density, potential temperature and salinity; $\mathbf{F}$ represents the viscous force term and $Q$ denotes the diffusion of temperature (by subscript `$T$') and salinity (by subscript `$S$'). 
% %The southern boundary is defined at $30.5^\circ$N.

%The Coriolis parameter is approximated as $f=f_0+\beta y$, where, 
%$f_0=7.3\times10^{-5}$ s$^{-1}$ and $\beta=2.1 \times 10^{-11}$ m$^{-1}$ s$^{-1}$. The term $f_0$ 
%is defined at the southern boundary of the domain, which is $30.5^\circ$N. 

We intend to simulate the Mediterranean sea, the horizontal domain extent of which is $8\degree$W - $36\degree$E in longitude and $30.5\degree$N - $46\degree$N in latitude. We consider a grid resolution of  $\sim$ $0.1\degree \times 0.1\degree$, which results in  $435\times140$ grid points. % in the longitude direction and $140$ in the latitude direction, 
%which gives grid resolutions of $\sim$ $0.1\degree \times 0.1\degree$. 
In the  vertical (radial) direction we consider $60$ non-uniformly spaced grid points, which varies from $1$ m at the free surface to a maximum value of $200$ m in the deeper regions. The horizontal viscosity and diffusivity terms are modeled using bi-harmonic formulation with $1.5\times10^{10}$ m$^4$/s as both viscosity and diffusivity coefficients \citep{calafat2004comparsion}.  Following Wunsch and Ferrari\cite{wunch2004vertical}, the vertical eddy-diffusivity for temperature and salinity are considered to be $10^{-5}$ m$^2$s$^{-1}$. Likewise, the vertical viscosity coefficient is assumed to be $1.5 \times 10^{-4}$ m$^2$s$^{-1}$, following  Calafat et al. \cite{calafat2004comparsion}. The lateral and bottom boundaries satisfy no-slip and impenetrability conditions. The numerical model incorporates implicit free surface with partial-step topography formulation \citep{adcroft1997representation}. 

The bottom topography of the Mediterranean sea (see Fig.\ \ref{fig:sea}(a))  is taken from The General Bathymetric Chart of the Oceans' (GEBCO) gridded bathymetric datasets \citep{weatherall2015new}. 
%The topography ranges from $8\degree$W - $36\degree$E in longitude and $30.5\degree$N - $46\degree$N in  latitude. 
The currently available resolution, based on ship-based survey and satellite altimetry combined, is $30$ arc-seconds. For our numerical simulation purposes, the topography data has 
been interpolated to our grid resolution.
%The amplitude of the topography varies from $0$ to $\sim$ 5000 m.  

% \begin{figure}
% \centering
% \includegraphics[width=1.0\linewidth]{topo_mitgcm_1}
% \caption{Mediterranean sea bathymetry $(h(\theta, \phi))$ reconstruction using MITgcm. (A)  Actual bathymetry obtained from GEBCO, and (B) reconstructed bathymetry. The color contours represent  
% depth $h$ (in m).
% %negative sign implying that the datum is at the sea-surface.
% }
% \label{fig:sea}
% \end{figure}

The numerical model has been initialized with 3D temperature, salinity, horizontal velocity (both zonal and meridional components), and free surface elevation data from Nucleus for European Modelling of the Ocean (NEMO)
re-analysis data obtained from Copernicus Marine Service Products \citep{Karina2016copernicus}. The input variables, taken on  12$^{\mathrm{th}}$ December 2017, are time-averaged (over that given day), and then interpolated to the grid resolution. 
% The wind-stress is obtained from the $6$-hour European Centre for Medium-Range Weather Forecasts (ECMWF) ERA-Interim re-analysis wind velocity data $(\mathbf{u}_a)$ at $10$ m above the sea level \citep{dee2011era}. The wind-stress is calculated using Eq.\ (\ref{eq:for_wind_stress}), in which the drag coefficient, $C_d$, is calculated for every $6$ hours as a function of wind velocities and temperature differences between air ($T_a$) and sea surface ($T_s$) using the following polynomial formula \citep{hellerman1983normal}:
% \begin{equation}
% \label{eqn22}
% C_d = \alpha_1 + \alpha_2|\mathbf {u}_{a}| + \alpha_3(T_a-T_s)+\alpha_4|\mathbf {u}_{a}|^2 
%       \\
%       + \alpha_5(T_a-T_s)^2+\alpha_6|\mathbf {u}_{a}|(T_a-T_s).
% \end{equation}
% Here $\alpha$ with subscripts $1, 2, ... 6$ are constants, the values of which are
% taken from \citet[Eq.\ (11)]{hellerman1983normal}.
% These data have been taken on the same date as the initialization data for the numerical model
% from Copernicus Marine Service Products (12$^{\mathrm{th}}$ December 2017).  $T_a$ is taken at $2$ m %\todo{temperature taken from 2m} 
% above the sea level, and is obtained from the ECMWF ERA-Interim re-analysis data  on the same date. Likewise, $T_s$  is obtained  from NEMO-MED re-analysis data of Copernicus Marine Service Products. 
The model has been integrated for $30$ days with a constant time-step of $100$\,s so as to reach a quasi-steady state.
%For computational efficiency, the parallel algorithm of the code has been exploited. The computation has been performed with a $4$-core Intel$\textsuperscript{\textregistered}$ Xeon processor, the computational time being  $\approx 120$ CPU-hours.

For calculating $\eta_g$,  the free surface velocity over the last $7$-days of the simulation are taken and subsequently time-averaged, yielding the geostrophic velocity.  At boundaries we set $\eta_g=0$ and solve Eq.\ (\ref{eq_geo_tilt}). The geostrophic velocity satisfies the horizontal divergence-free condition, hence  contains no information about the the bottom topography. Topography information is contained in the ageostrophic velocity part.

In order to do the reconstruction, we have averaged the free surface velocity field over $12$ hours. This averaging time has been judiciously chosen -- not too long so that the flow is geostrophic, and not too short so that the surface elevation gets affected by surface waves.  
%For bathymetry reconstruction we solve the spherical coordinate version of  Eq.\ (\ref{SWC2}), where the free surface elevation and velocity field are  {$12$-hours time-averaged}. 
For $H$ we have taken a resolution of $\sim 0.5\degree$ in both latitude and longitude directions so as to mimic the large scale topographic structure. For the reconstruction we solve the spherical coordinate version of  Eq.\ (\ref{SWC2}). For ease of understanding, the solution algorithm is given below:
{
\begin{algorithm}[H]
\caption{ Procedure for finding $b$}
\label{alg:the_alg}
\begin{algorithmic}[1]
\Procedure {inverse bathymetry}{}
\State input: $H$, $u$ and $v$ at the free-surface ($\eta$)
\State process:
\State Step 1: Find the geostrophic flow induced tilt -- take 7 days time average of $u$ and $v$ to get the geostrophic flow and use Eq.\ (\ref{eq_geo_tilt}) to find $\eta_g$.
\State Step 2: Find  $\eta_b$ -- take 12 hours time average of $\eta$ and subtract $\eta_g$ to get $\eta_b$ (use Eq.\ (\ref{eq:eta_avg})).
\State Step 3: Solve the spherical coordinate version of Eq.\ (\ref{SWC2}) to get $b$. 
\EndProcedure
\end{algorithmic}
\end{algorithm}
}

%, where the free surface elevation induced by 
%bathymetry and averaged-velocity field are used. The equation has been discretized using central finite-difference method and solved iteratively using tridiagonal-matrix algorithm. For, boundary conditions, we have used $b=0$ at the boundaries.} 
The reconstructed bottom topography, shown in  Fig.\ \ref{fig:sea}(b),  is   $\approx 97.6\%$ accurate.
%  $\approx 98.3\%$
%We emphasize here that the spherical coordinate version of  Eq.\ (\ref{SWC2}), used for bathymetry reconstruction, is a \textit{diagnostic equation} since we have not imposed shallow water approximation anywhere in MITgcm. Hence the large-scale 2D flow is primarily important for bathymetry reconstruction, additional effects of density stratification and three-dimensionality are insignificant.
%  $\approx 98.3\%$
We emphasize here that the spherical coordinate version of  Eq.\ (\ref{SWC2}), used for bathymetry reconstruction, is a \textit{diagnostic equation} since we have \emph{not} imposed shallow water approximation anywhere in MITgcm. 
Hence the large-scale 2D flow is primarily important for bathymetry reconstruction, additional effects of density stratification and three-dimensionality are insignificant.
%{The shallow water model has been applied to Mediterranean sea to study the dynamics of
%sea surface variation due to tidal flow\citep{molines1991modeling} and also has been used to study propagation of the tsunami wave\citep{samaras2015simulation}.}

As mentioned earlier, Fourier transform of the free surface provides an alternative technique to bathymetry reconstruction. Fig.~\ref{fig:sea}(c) shows the Fourier transform of the free-surface anomaly (after removing the geostrophic flow induced tilt) of the boxed region (red-colored line) marked in the Mediterranean sea (see Fig.~\ref{fig:sea}(a)). The free surface contains stationary features (which contains the information about the underlying bathymetry), marked by `SF', and wave-like signatures, marked by `PW'. By inverting SF, one can reconstruct the bathymetry of the boxed region.

{
Finally we attempt to reconstruct ocean bathymetry completely from 
re-analysis data.
%observation data.   
 We have first chosen Red sea in this regard, the necessary data for which is obtained from  HYCOM (Hybrid Coordinates Ocean Model) based NOAA Global forecast system \citep{chassignet2007hycom}. It provides $3$-hourly global ocean data with a horizontal resolution of $1/12\degree$ for $40$ vertical depth levels. 
%  HYCOM assimilates real-time satellite (Envisat, GFO and Jason-1) data, in-situ measurements of sea surface height, sea surface temperature, 3D temperature and salinity fields (Argo, CTDs and moorings),  as well as  Geostationary Operational Environmental Satellite (GEOS) data \citep{chassignet2009us}. 
% Furthermore, 
 The model  uses \textit{ETOPO5} topography data of  $1/12\degree$ resolution  \citep{noaadigital}. We have taken $5$ datasets of 2017, each of $7$-day length: $5^{\mathrm{th}}-11^{\mathrm{th}}$ March, $12^{\mathrm{th}}-18^{\mathrm{th}}$ April, $8^{\mathrm{th}}-14^{\mathrm{th}}$ May, $15^{\mathrm{th}}-21^{\mathrm{st}}$ June and $9^{\mathrm{th}}-15^{\mathrm{th}}$ July.
 Corresponding to each dataset, first the geostrophic velocity is calculated by performing a $7$-day time-average and calculate $\eta_g$ using Eq.\ (\ref{eq_geo_tilt}). In oceans, wind-stress $\bm{\tau}$ is always present, and is obtained from the wind velocity data \cite{gill1982atmosphere}:
 %; see equation (\ref{eqn21}):
 \begin{equation}
 \label{eq:windwind}
\centering
\boldsymbol{\tau} = \rho_a C_d \mathbf{u}_{a} |\mathbf {u}_{a}|, 
\end{equation}
where $\rho_a=1.2$ kg/m$^3$ is the density of air, $C_d$ is the drag coefficient and $\mathbf{u}_{a}$ is the wind velocity. The value of $C_d$ is calculated for every $6$ hours as a function of wind velocities and temperature differences between air ($T_a$) and sea surface ($T_s$) using the following polynomial formula \cite{hellerman1983normal}:
\begin{multline*}
%\label{eqn22}
C_d = \alpha_1 + \alpha_2|\mathbf {u}_{a}| + \alpha_3(T_a-T_s)+\alpha_4|\mathbf {u}_{a}|^2 
         \\+ \alpha_5(T_a-T_s)^2+\alpha_6|\mathbf {u}_{a}|(T_a-T_s),
\end{multline*}
where $\alpha$ with subscripts $1, 2, ... 6$ are constants, the values of which are
taken from Eq.\ $(11)$ of Hellerman and Rosenstein \cite{hellerman1983normal}. $T_a$ is taken at $2$ m %\todo{temperature taken from 2m} 
above the sea level, and is obtained from the ECMWF ERA-Interim re-analysis data  on the same dates of interest. Likewise, $T_s$  is obtained  from NEMO-MED reanalysis data of Copernicus Marine Service Products. Wind stress causes quasi-stationary free surface elevation $\eta_s$, which is given by\cite{janzen2002wind}:  
%The quasi-stationary free surface elevation $\eta_s$ caused by  wind-stress is given by \cite{janzen2002wind}: 
\begin{equation}
\nabla \eta_s =\frac{\bm{\tau}}{g\rho_w H},
\label{eq:wind_stress_main}
\end{equation}
 where $\rho_w$ is the density of water, and it is assumed that the wind-stress is small  and therefore does not affect the inertial acceleration. 
 %Here we outline a generic strategy for removing  free surface tilts induced by small wind-stress.
 Depending on whether we are using Cartesian or spherical coordinate system, the $\nabla$ operator is chosen accordingly.\\
 \indent In reality, wind-stress can occasionally become large (e.g.\ storm events), making the wind-stress induced tilt calculation invalid. For this reason, the datasets are  carefully selected such that low wind velocity is ensured. 
%  This surface wind-stress induce a free-surface slope and at steady-steady balance set quickly between surface-wind stress and free-surface slope. Mathematically, it can be expressed as \citep{menemenlis2007atlantic} $$\nabla \eta_s =\frac{\bm{\tau}}{g\rho_w H}$$ where $\eta_s$ is the free-surface elevation due to wind-stress, $\rho_w$ is density of the water. Therefore, Eq.\ (\ref{eq:eta_avg}) can be recast as
% \begin{equation}
% \langle \eta \rangle=\eta_g+\eta_s+\eta_b,
% \label{eq:eta_avg1}
% \end{equation} 
The time-averaged free surface elevation is now given by
\begin{equation}
\langle \eta \rangle=\eta_g+\eta_s+\eta_b,
\label{eq:eta_avg1}
\end{equation} 
and, although it is more complicated than Eq.\ (\ref{eq:eta_avg}), still we have the recipe of removing $\eta_s$ following Eq.\ (\ref{eq:wind_stress_main}). After removing both $\eta_g$ and $\eta_s$, only free surface feature left is $\eta_b$.
%However, reconstruction equation does not change which is same as Eq.\ (\ref{SWC2}), although in spherical coordinate system.
At last, the bathymetry is reconstructed using the spherical coordinate version of Eq.\ (\ref{SWC2}), in which  the free surface velocity and elevation data are $12$-hours time-averaged. The  resolution of the mean depth $H$ is taken to be $6$ times coarser ($1/2 \degree$). For each dataset we obtain an inverted bathymetry map, the final map is the average of the five datasets. The original and the reconstructed bathymetry are compared in Figs.\ \ref{fig:red_sea}(a)-\ref{fig:red_sea}(b); the average reconstruction error is  $12.51\%$. \\
\indent A  similar technique can be followed in reconstructing any other bathymetry. For example, we reconstruct Mediterranean sea bathymetry using the following $5$ datasets: $1^{\mathrm{st}}-7^{\mathrm{th}}$ May, $12^{\mathrm{th}}-18^{\mathrm{th}}$ June, $7^{\mathrm{th}}-13^{\mathrm{th}}$ July, $20^{\mathrm{th}}-26^{\mathrm{th}}$ August and $15^{\mathrm{th}}-21^{\mathrm{st}}$ September. The actual and reconstructed bathymetries are shown in Figs.\ \ref{fig:red_sea}(c)-\ref{fig:red_sea}(d), the average reconstruction error is $12.68\%$.
\begin{figure}
\centering
\includegraphics[width=1.0\linewidth]{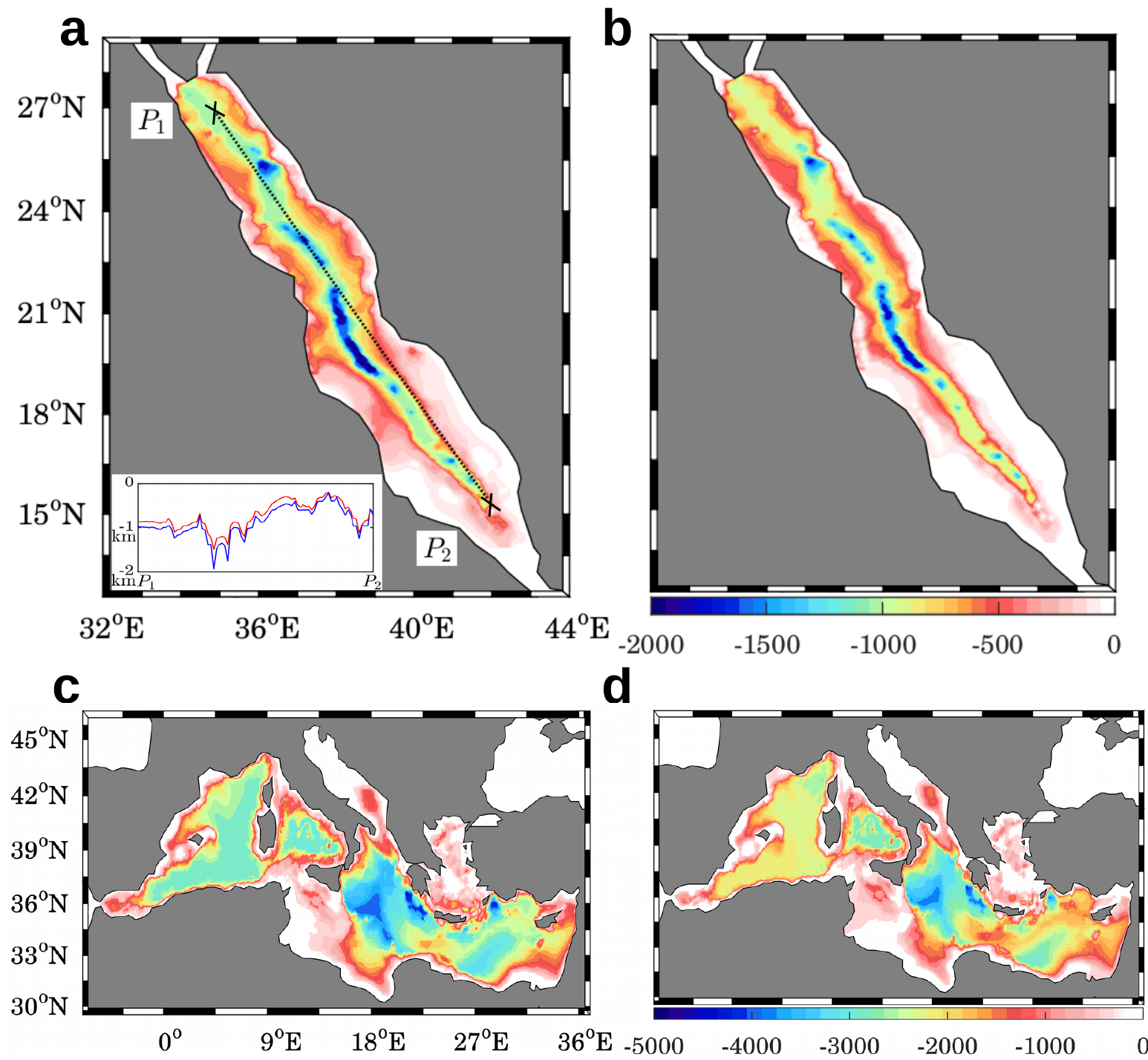}
\caption{{Bathymetry reconstruction from real data. (a) Original and (b) reconstructed Red sea bathymetry. 
The inset of (a) shows a comparison between the original (blue line) and reconstructed (red line) topography along the line $P_1P_2$ ($14.31\%$ error).  
(c) Original and (d) reconstructed Mediterranean  sea bathymetry.  The color contours in (a-d) represent  depth $h$ (in m) from the free surface.}}
\label{fig:red_sea}
\end{figure}}
In conclusion, we have shown that for shallow, free surface flows, the geometric information of the underlying topography remains embedded in the free surface. Based on the shallow water mass conservation equation, we have proposed a simple inversion technique that successfully reconstructs the bottom topography  from the free surface elevation and velocity field. 
{ We have applied this technique to (i) a toy ocean model, (ii) global circulation model (MITgcm) initialized by re-analysis data, and finally, (iii) purely re-analysis data.  For the MITgcm case, we reconstruct Mediterranean sea bathymetry  of $0.1\degree$ resolution with  $97.6$\% accuracy. For pure re-analysis data, both Red and Mediterranean sea bathymetries of $1/12\degree$ resolution are  reconstructed with  $\approx 90$\% accuracy.}

%Furthermore, we applied this technique  to  reconstruct the Mediterranean sea bathymetry  of $0.1\degree$ resolution with  $97.6$\% accuracy. 
%{Although, the simulation has been carried out through a controlled environment by supplying time-instant input data.}  
%Here, the Mediterranean sea was simulated using MITgcm, which was initialized with contraoled re-analysis data. 

%We  show that uniform resolution bathymetry map with $\approx 90$\% accuracy can be reconstructed from the ocean free surface elevation and velocity data obtained via satellite altimetry.
%Since our technique is not based on gravity anomaly, it can easily resolve long wavelengths ($>160$ km), which is not possible in altimetric bathymetry due to isostatic compensation. 

In conjunction with ship echo-soundings, our reconstruction technique may provide a highly accurate global bathymetry map in the future. 
% However, the proposed theory remains to be tested in real-ocean scenarios.  Wind-stress can apparently be of some concern since it can give rise to stationary features on the free surface. Such features must be removed for correct bathymetry reconstruction. By inspecting the available wind data, one can choose any appropriate 12-hour period during which the wind is relatively calm, and hence circumvent this issue. 
%using satellite altimetry, which provides real-time free-surface height and surface velocity field.
{ The problem remains to be tested on data fully obtained from satellite altimetry.} At present,  satellites do not provide very reliable information in the horizontal-scale of $\lesssim 100$ km.
%, where the interest of this work lies. 
In near future, the Surface Water Ocean Topography (SWOT) satellite mission will revolutionize the field by providing information at unprecedented scales of $15-25$ km, which is of an order of magnitude higher resolution than that of current satellites \cite{gaultier2016challenge}. Our technique will be specifically useful in obtaining accurate bathymetry maps of the shallow coastal regions, where the estimated reconstruction time by ship-based surveying is 750 ship-years. 
%Additionally, our technique may be applied to reconstruct the bathymetry of  numerous remote and virtually uncharted regions, especially those in the Antarctic and the Arctic. 
%Moreover, the procedure being quite general, it can be applied to reveal the surface of planets covered with thin, opaque atmosphere.  

This work has been partially supported by the following grants: IITK/ME/2014338, STC/ME/2016176 and
ECR/2016/001493.
  \bibliography{pnas-sample}
\end{document}